\documentclass{PoS}
\usepackage{subfig}
\usepackage{float}

\title{Solving the Dirac equation on QPACE}

\ShortTitle{Solving the Dirac equation on QPACE}

\author{\speaker{Andrea Nobile}\\%\thanks{A footnote may follow.}\\
        Institute f\"ur Theoretische Physik, Universit\"at Regensburg, 93040 Regensburg, Germany\\
        E-mail: \email{andrea.nobile@physik.uni-regensburg.de}}

\abstract{We discuss the implementation and optimization
challenges for a Wilson-Dirac solver with Clover term
on QPACE, a parallel machine based on Cell processors
and a torus network.\
We choose the mixed-precision Schwarz preconditioned FGCR algorithm in order to circumvent network bandwidth and latency constraints, to make efficient use of the multicore parallelism and on-chip memory, and to achieve flexibility in the choice of lattice sizes.
We present benchmarks on up to 256
QPACE nodes showing an aggregate sustained performance of about 10 TFlops for the complete solver and very good scaling. 
}

\FullConference{The XXVIII International Symposium on Lattice Field Theory, Lattice2010\\
		June 14-19, 2010\\
		Villasimius, Italy}

\begin{document}

\section{Introduction}
The simulation of lattice QCD including dynamical fermions is computationally challenging mainly due to the 
need of performing lattice Dirac operator inversions.
These inversions are needed within the HMC algorithm to compute the contribution to the force which arises from the fermionic determinant.
%Due to the relatively simple structure of the most computationally intensive algorithms involved in the simulation of lattice QCD, there is a long history of custom machines, recent examples are \cite{apenext} and \cite{qcdoc}. 
%The advantage of a special purpose machine is the superior efficiency, respect to a general purpose machine, in executing the particular tasks it is designed for.
Many tasks involved in the simulation of lattice QCD show an high degree of parallelism at different levels.
At the lattice level the natural parallelization strategy consists of subdividing the lattice into sub-lattices, each sub-lattice is then associated to a different processing node of the machine. In order to exploit efficiently the intrinsic parallelism, a low-latency, high-bandwidth, scalable network is needed to connect the different processing nodes.

QPACE \cite{Baier:2009yq} is the latest custom designed massively parallel supercomputer for LQCD applications.
The architecture consists of a 3D torus of processing nodes based on a commodity processor, the IBM  PowerXCell 8i.
The nodes are interconnected by a custom network which enables low-latency and high-bandwidth nearest neighbor communications. 
Each node is thus equipped with six links to connect to the six neighbors in the 3D torus mesh. \\
While the floating point (FP) performance of the processors has seen a continuous increase in the past years, the latency of the memory accesses and network has not improved at a similar scale. This problem can be mitigated by increasing the amount of on-chip memory.
Making optimal use of this on-chip memory has become crucial for the optimization of numerical software.

\section{Cell processor}
%The Cell processor is in many ways a revolutionary design. In order to achieve a high FP peak performance and to cope with the long latencies of the memory accesses, designers have chosen to simplify some aspects of the architecture.
The Cell architecture consists of nine cores: one Power Processing Unit (PPU) able to run generic code such as the operating system, and eight Synergistic Processing Units (SPU) for which a simple architecture was chosen, optimized for FP intensive applications.
%The nine cores, the memory controller, and the I/O interface are connected by an on-chip ring bus (EIB). 
The Cell architecture is described in \cite{cbea}. 
Among the different design choices done on the SPU to reduce core size, maximize FP performance, and allow latency hiding, the most radical one is the explicit separation of two levels of the memory hierarchy.
While in conventional processors the fact that there is a small amount of fast memory and a large amount of slow memory is completely transparent to the programmer, being the hardware responsible to move data between the memory spaces, on the Cell, this responsibility is completely left to the programmer, who needs to issue \emph{direct memory access} (DMA) commands in order to move data. 
%The explicit memory hierarchy handling makes the programming very similar to assembly in which processor's registers are exposed and instructions are issued to move data between memory and registers. 
This particular feature makes programming a complex, time consuming and error prone task even for routines which do not need a particular high level of optimization.
On the other hand, by allowing full control of the fast memory, it is at least in principle possible to make an optimal use of the resource.

The instruction set of the SPU is similar to Intel SSE. Almost all the instructions operate on vector types with operands and results being stored in 128 general purpose, 128-bit wide registers. Peak performance is obtained by using the fused multiply-add instruction.
It is mandatory to use vector instructions in all the cases in which reasonable performance is needed.

%The  integer instructions operate also on vector types. There are no scalar registers, this means that scalar operations, both on floating point and integer values, are particularly slow due to the overhead of the necessary shuffle instructions. 
%It is thus mandatory to use vector instructions in all the cases in which reasonable performance is needed. This is accomplished by restructuring the code and data structures in order to use vector intrinsics within C code. Given the overhead associated with scalar operations, in extreme cases, the vectorization of some integer operations like pointer manipulation can be beneficial.

Access to main memory from the SPU is subject to alignment constraints. The minimum allowed alignment is 16B, optimal memory bandwidth is achieved with 128B aligned DMA where the size is a multiple of 128B. Close to peak bandwidth is obtained already for a DMA size of 128B.
Given the high FP performance of the chip (200 GF single precison and  100 GF double precision) many tasks are limited by memory bandwidth (8B/cycle for the whole Cell), it is thus very important to access memory in the most efficient way.

\section{QPACE torus network}
The QPACE torus network is capable of exchanging messages between nearest-neighbor nodes in the 3D torus. The \emph{send} operation corresponds to a DMA put operation on the address range of the network links. At the receiving side the network processor forwards the data to the SPU's local stores or to main memory to a user-specified address. After all the data corresponding to a receive operation has been forwarded, a particular memory location is updated to \emph{notify} the completion of the receive operation. The QPACE torus network supports eight \emph{channels}. Up to eight messages can travel concurrently on the same link. This feature enables the concurrent communication between SPUs of different nodes. The network supports also \emph{remote offsets}.%, by adding an offset to destination address of the put DMA, the data
%are forwarded to the destination address specified by the receive operation augmented by the offset. 
This feature can be used to merge different receive operations in a single one, when different SPUs have to send data to different locations of the same buffer on the receiving node. %Messages have to satisfy alignment constraints: send and receive buffers must be 128B aligned and the message size must be a multiple of 128B. 

\section{Algorithm Choice}
The conjugate gradient (CG) is one of the most popular solvers used for lattice QCD simulations. CG gained its popularity due to good convergence properties, simplicity, and robustness. The most numerical intensive task in the CG algorithm is the matrix-vector product which consists in the application of the lattice Dirac operator to a spinor field. Almost all of the optimization effort is usually spent on this particular task.
As was shown in \cite{phd}\cite{lat07} the performance of an optimal implementation of the Wilson-Dirac operator without even-odd preconditioning on the Cell processor is bounded by the memory bandwidth to $\epsilon_{fp}=34\%$. This estimate relies on the assumption that it is possible to reach full memory bandwidth. A benchmark of this implementation shows a performance $\epsilon_{fp}=24\%$. This optimal implementation is subject to a lattice size constraint. The maximum local lattice size for a single precision computation is limited to $L_0\times 10^3$ and to $L_0\times 8^3$ for single precision where $L_0$, the time extent of the lattice, is not constrained. This limitation comes from the size of the local store.
This optimal implementation is also able to tolerate a network latency of a few $\mu$s, i.e. O(10,000) clock cycles.
In the case of an even-odd preconditioned Wilson-Dirac operator, both memory and network accesses become more problematic. First we notice that during the application of the operator, the spinor field on the borders of the local lattice must be communicated two times, since this operator couples next to nearest neighbor lattice points. A feasible implementation of this operator requires two sweeps of the lattice each of which requires access to all the SU(3) link variables while doing half of the needed floating point operations.
The memory bandwidth limitation thus becomes more critical and the estimated performance, taking into account the efficiency of the memory goes below $20\%$, even without considering the network.

In order to circumvent these constraints, including the limited flexibility on the choice of the local lattice sizes, we consider the SAP-GCR algorithm that was proposed by L\"uscher \cite{SchwarzII}.
The Schwarz Alternating Procedure (SAP) belongs to the class of domain decomposition methods. The lattice is divided into non-overlapping blocks which are chessboard-colored. One iteration (cycle) of the SAP proceeds as follows: the Wilson-Dirac equation first is solved on all the \emph{white} blocks, the residue is then updated both on the \emph{white} blocks, and on the internal border of the \emph{black} blocks, then the equation is solved on the \emph{black} blocks, with the source consisting in the updated residue, and finally the residue is updated on the \emph{black} blocks and on the internal border of the \emph{white} blocks.
The blocks are solved with a fixed number of iterations using a simple iterative solver (MR) and Dirichlet boundary conditions.
In this procedure the only step which requires network communications is the update of the residue on neighboring blocks. Since the updated residue is needed only after half of the blocks are solved, it is possible to completely overlap communication and computation. The procedure is thus able to tolerate very high network latencies without performance losses. The network bandwidth requirement is a fraction $1/(n_{it}+1)$ of the 
requirement of the equivalent Wilson-Dirac operator on the whole lattice, where $n_{it}$ is the number of iterations used in the block solver.
The SAP procedure is used as preconditioner inside a FGCR iteration. The outer FGCR iteration is able to tolerate variations in the preconditioner since the preconditioner is obtained with an iterative method and thus is not a stationary operator. FGCR is mathematically equivalent to FGMRES \cite{saad} and requires to store two vectors per iteration. This requirement translates in the practical need of restarting the FGCR recursion. Mixed precision is obtained through iterative refinement \cite{itref}. All the operations can proceed in single precision, only when the FGCR recursion is restarted, the residue is computed in full double precision, and the double precision solution is updated by adding the single precision correction.  
The most time consuming task in the SAP is the solution of the Wilson-Dirac equation on the blocks. Since the block size can be chosen such that all the data necessary for a block solve fit the local store, SAP is able to achieve a high sustained performance by making an efficient use of the local store. 
The local lattice sizes are only constrained to be multiples of the block sizes. Table \ref{table:comp} summarizes the comparison between the CG and SAP-GCR. 

\begin{table}[ht]
\caption{Comparison between CG and SAP-GCR on QPACE}
\centering
\begin{tabular}{lccc}
       & CG         & SAP-GCR  \\
    \hline
    Lattice size flexibility        &   poor    & reasonable \\
    E/O preconditioning         &    perf. penalty     & yes (block solver) \\
    Network bandwidth req.         &   high             & low  \\
    Network latency tolerance         &   moderate       & high  \\
    Memory bandwidth req.   &  high & moderate \\
    
\end{tabular}
\label{table:comp}
\end{table}

\begin{figure}[t]
\centering

\includegraphics[width=0.6\textwidth]{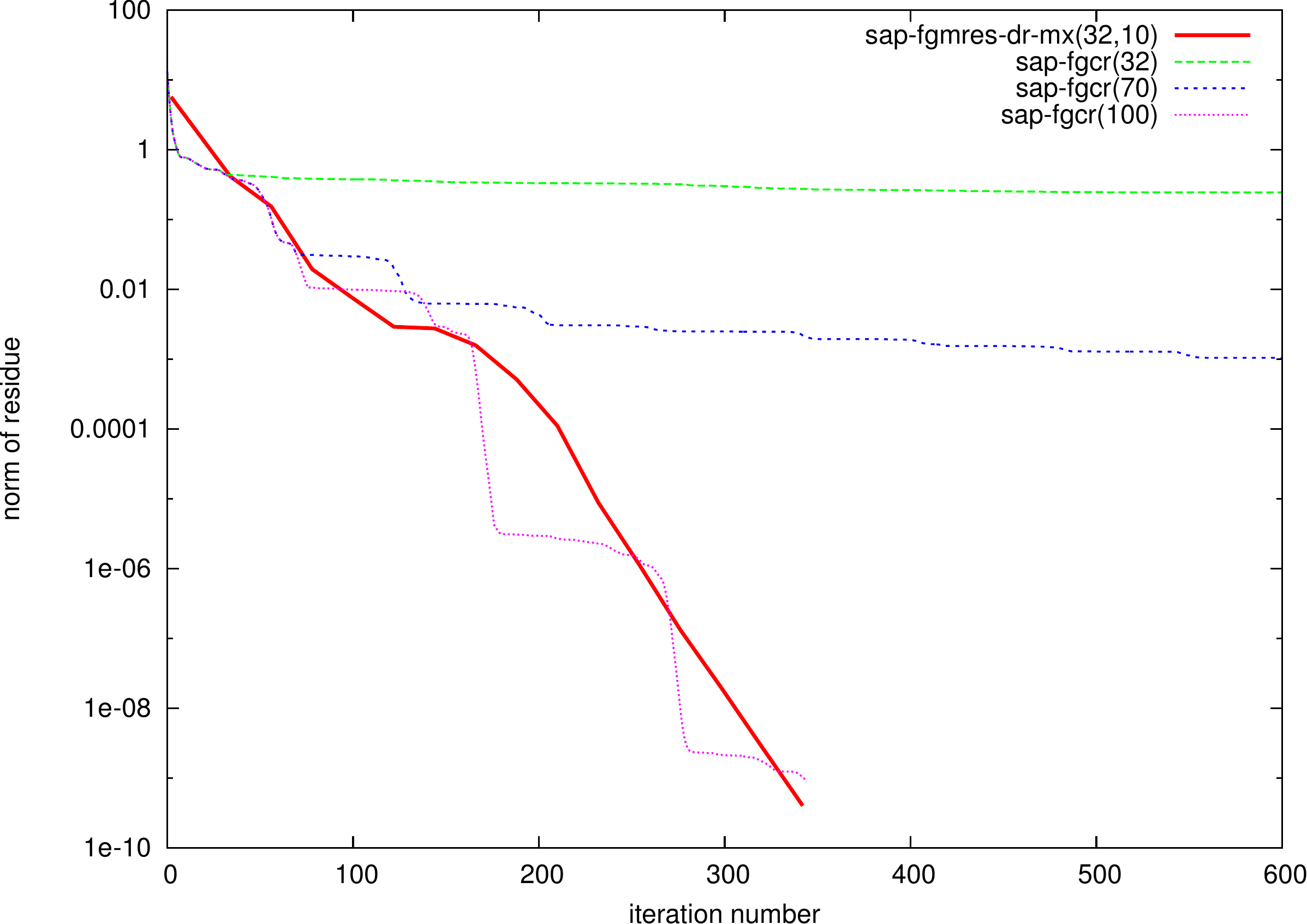}
\caption{Convergence of SAP-GCR with Krylov subspace dimension 32, 70 and 100, SAP-FGMRES-DR with Krylov subspace dimension 32 including 10 deflated vectors. $16^3\times32$, $\beta=5.29$, $\kappa_{sea}=0.13500$, $\kappa_{val}=0.13768$, $\kappa^{c}_{val} \approx 0.13770$}
\label{sapfgmresdr}

\end{figure}

A drawback of the SAP-GCR algorithm is that convergence is not guaranteed. When the FGCR recursion is restarted, the information contained in the Krylov subspace is lost. This can result in a reduction of the convergence rate of the algorithm or, in the worst case, stagnation. In order to avoid this problem we introduced deflation in the form of FGMRES-DR and we eventually foresee automatic switching to CG for the most difficult cases. The idea of \emph{deflated restarts} consists in keeping some information between restarts by computing approximate eigenvectors and including them in the Krylov supspace for the subsequent restart. At the end of a FGMRES cycle of lenght $m$, $m$ \emph{harmonic Ritz} pairs are computed using the small matrix $H$ built during the Arnoldi steps. The harmonic Ritz vectors corresponding to the smallest $k < m$ approximate eigenvectors are used to build an orthonormal basis for the deflation subspace. This subspace, augmented with the residue, becomes the initial Krylov subspace for the subsquent restart. Figure \ref{sapfgmresdr} shows the convergence as a function of the Krylov subspace dimension for SAP-GCR and a comparison with the mixed precision SAP-FGMRES-DR \cite{sapfgmresdrp} for $\kappa$ close to $\kappa_c$.  SAP-FGMRES-DR is able to converge in many cases in which SAP-GCR fails, it achieves the same convergence rate of SAP-GCR with much less vectors resulting in less critical memory requirements and faster execution. FGMRES-DR is the flexible generalization of GMRES-DR \cite{gmresdr}. 

\section{SAP implementation}
The most time consuming task in the SAP-GCR algorithm is the Schwarz Alternating Procedure. Most of the floating point operations are spent inside the block solver routines. It is thus natural to focus the optimization efforts on these tasks. 
Given the memory access alignment constraints, the user-controlled local store and the need of making efficient use of the SIMD instructions, it is important to choose a data layout that maximizes performance and at the same time simplifies the programming. Lattice points are ordered such that points belonging to the same block are contiguous. Inside the blocks, lattice points are divided into two sets, even and odd. In this way DMA operations performed to move block fields between main memory and local store are greatly simplified and the performance is maximized.
For the spinors layout, different possibilities were analyzed by counting the floating point and shuffle instructions involved in the application of the Wilson-Dirac operator for the different layouts. The final choice is such that the indexes, from the slowest running to the fastest are: \emph{color, spinor, complex}. In this way, when a spinor is loaded into 4-way SIMD registers, each register contains components of the same color.
The SU(3) matrices, consisting in 72B arrays in single precision, do not satisfy the basic 16B alignment constraint coming from the size of the registers, we have thus chosen to pad the SU(3) structure to 80B. While this choice introduces an overhead by wasting a fraction of the memory bandwidth, it greatly simplifies the code by avoiding the otherwise necessary and slow manual alignment operations.
The multicore parallelism is exploited by parallelizing the block solver on the eight cores. The block is divided along the time dimension among the cores, thus constraining the block size in the time direction to 8 (the number of cores).
Each core works thus on a time-slice of the block.
Complete overlap of memory accesses with computation can be in principle done given a sufficient local store size.
We have chosen to have the maximum possible block size to maximize the block solver performance, this limits the amount of data that can be prefetched.
The parallelization of the SAP among the QPACE nodes is not straightforward but the structure of the algorithm fits nicely the features of the network.
 After one block is solved, the data necessary for the update of the residue on neighboring blocks residing on remote nodes is available and is sent via a DMA put to the remote nodes directly from the SPUs. While it is in principle necessary to distinguish between remote and local blocks (data for remote blocks must be sent through the network), this distinction is not done at the level of the SPU code resulting in a further simplification of the SPU code. The PPU control code passes the addresses of the network links to the SPUs for the remote blocks while passes memory addresses in the case of local blocks. All the complexity associated with the network is handled by the PPU control code which is also responsible for issuing receive commands to the network processor.
In order to distinguish between data sent by different SPUs simultaneously, the \emph{remote offset} feature is exploited such that all the data belonging to a given block border sent by the 8 different SPUs is received in a contiguous buffer with a single receive operation, reducing the overhead associated with receive commands. The channel feature of the network is exploited by having up to 8 concurrent blocks per link. Network send and receive operations are asynchronous allowing overlap between communication and computation.
% For each block which have faces on the border a 64KB memory page per face is allocated in order to allow data forwarding by the network processor.

\section{Performance}
The block solver performance is not limited by main memory or network. For the block solver alone,  we were able to obtain an impressive $\epsilon_{fp} = 50\%$ for a  block size of $8^2\times 6^2$, using the IBM xlc compiler. The compiler optimization flags used  for this benchmark cause deep code transformations and the resulting SPU executable tends to be very big. In production SPU executables, for which there is the need of additional subroutines and buffers for the complete SAP, we had to reduce the block size and preferred to use the GCC compiler which produces more compact code. We measured efficiencies $\epsilon_{fp} = 36\%$ , $\epsilon_{fp} = 34\%$ and $\epsilon_{fp} = 30\%$ for $8^2\times 4^2$, $8\times 2\times 6^2$ and $8\times 10\times 2^2$ block sizes respectively.
Performance is also affected by the overhead of the necessary shuffle instruction and DMA commands.
The measured performance for the complete SAP is  $\epsilon_{fp} = 25.9\%$ , $\epsilon_{fp} = 23\%$ and $\epsilon_{fp} = 19.3\%$ for $8^2\times 4^2$, $8\times 2\times 6^2$ and $8\times 10\times 2^2$ block sizes respectively. We used 4 iterations for the block solver and 10 SAP cycles.
In a typical HMC production run with a $32^3$x$64$ lattice at $\kappa=0.13632 , \beta=5.29$ on 256 QPACE nodes we found that $75\%$, $13\%$ and $12\%$ of the solver time is spent inside SAP, double precison Wilson-Dirac operator and spinor linear algebra, respectively.

\begin{figure}[t]
\centering
\subfloat[Scaling up to 64 nodes with $8\times 2\times 6^2$ and $8\times 10\times 2^2$ block sizes \label{scal1}]
{
\includegraphics[width=0.49\textwidth]{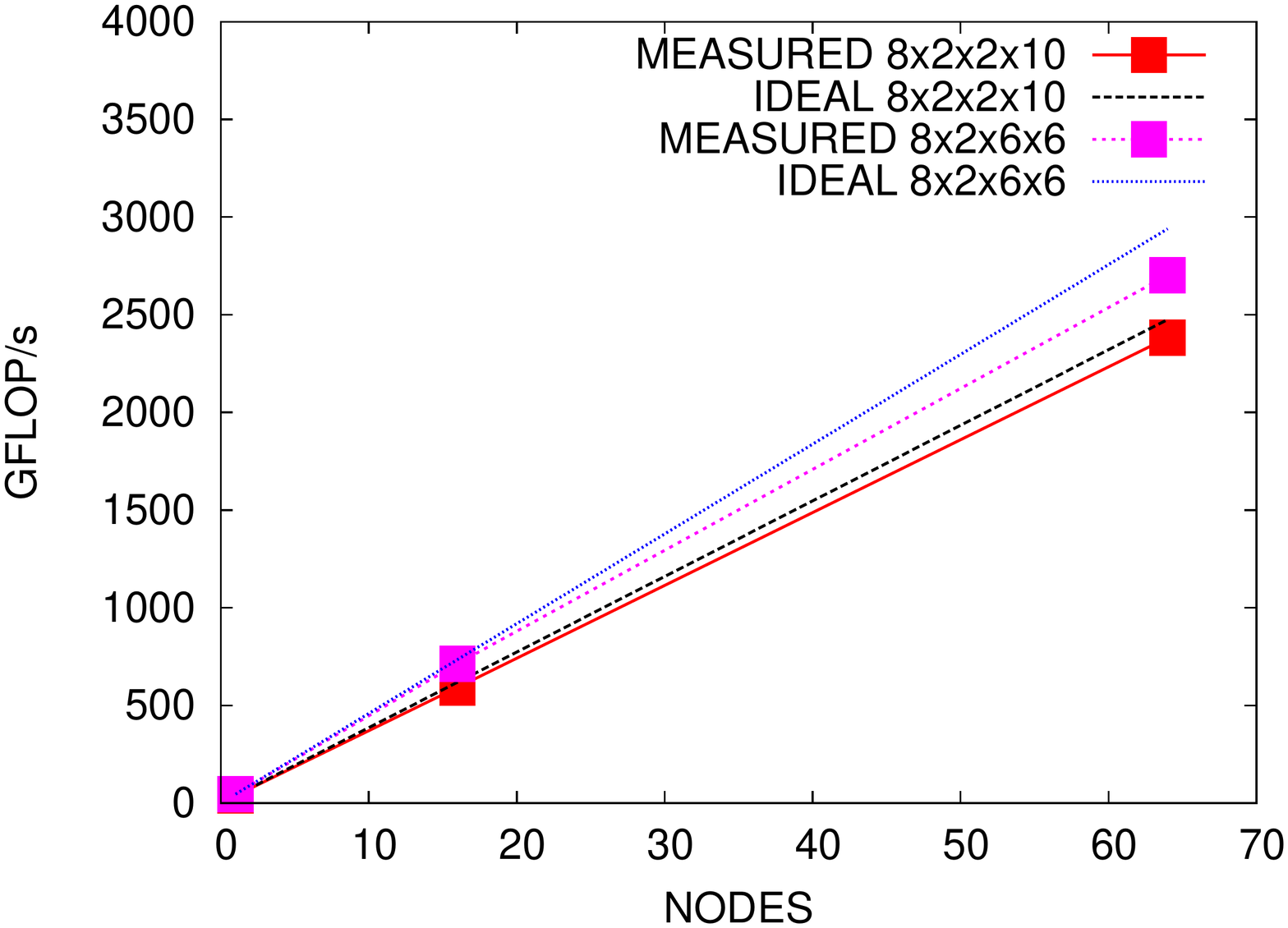}

}
\subfloat[Scaling up to 256 nodes with $8\times 4^2$ block size \label{scal2}]
{
\includegraphics[width=0.49\textwidth]{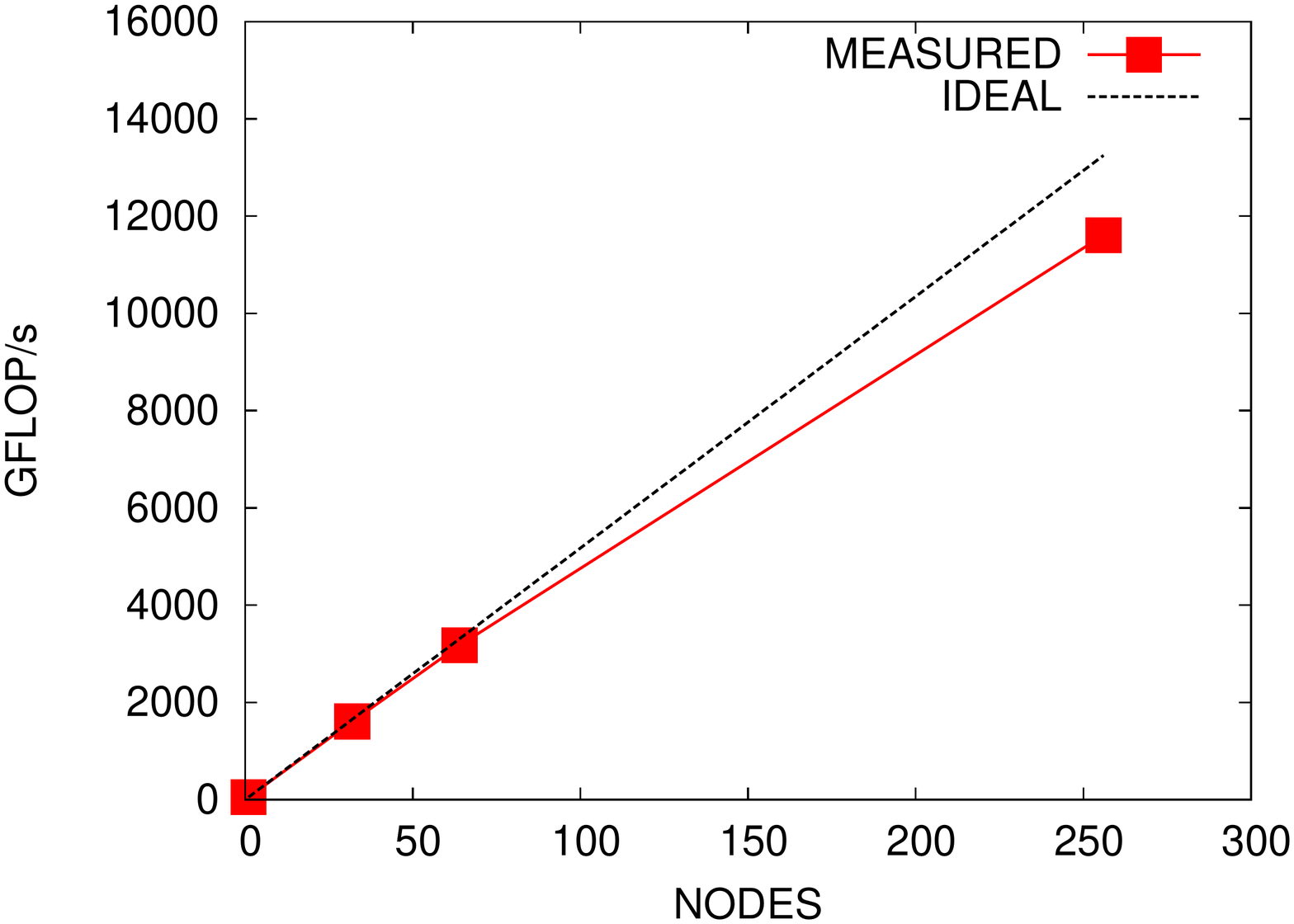}

}
\caption{SAP scaling}
\label{scal}
\end{figure}

In figure \ref{scal} we show the SAP performance and weak scaling for the three different block sizes used. The measured points are compared with the ideal scaling using the single node performance.
 
\section{Conclusions}
We demonstrated that the SAP-GCR and SAP-FGMRES-DR solvers fit nicely the processor, memory hierarchy and network architecture.
We showed the scalability of our implementation up to 256 nodes with very good performance.
Both solvers are currently integrated in BQCD \cite{bqcd} and are used for production runs on QPACE.

\section*{Acknowledgements}
We would like to thank D. Pleiter, H. Simma, T. Wettig, A. Frommer, Y. Nakamura, T.Streuer, S. Solbrig, B.Mendl and the whole QPACE team.
This work has been supported by the DFG (SFB/TR55, Hadron Physics from Lattice QCD).

\end{document}